# BRIDGE: A Model for Modern Software Development Process to Cater the Present Software Crisis

Ardhendu Mandal
*Lecturer, Department of Computer Science and Application, University of North Bengal*
*Raja Rammohanpur, PO-NBU, Dist-Darjeeling, West Bengal, Pin-734013, India.*
am.csa.nbu@gmail.com

*Abstract-* As hardware components are becoming cheaper and powerful day by day, the expected services from modern software are increasing like any thing. Developing such software has become extremely challenging. Not only the complexity, but also the developing of such software within the time constraints and budget has become the real challenge. Quality concern and maintainability are added flavour to the challenge. On stream, the requirements of the clients are changing so frequently that it has become extremely tough to manage these changes. More often, the clients are unhappy with the end product. Large, complex software projects are notoriously late to market, often exhibit quality problems, and don't always deliver on promised functionality. None of the existing models are helpful to cater the modern software crisis. Hence, a better modern software development process model to handle with the present software crisis is badly needed. This paper suggests a new software development process model, BRIDGE, to tackle present software crisis.

## I. Introduction

Now a day, computers running with special purpose application software are being used as an extensive aid to solve complex problems almost each and every place starting from gaming to engineering, industries applications, scientific research and different allied fields. These special purpose softwares are some times unique and distributed in nature with higher degree of complexity. Developing such complex software is not so easy because of the different constraints. Our existing software models do not provide adequate flexibility to be applied for such large and complex projects. So we must have a better software development process model that will help to overcome these challenges.

## II. Usage of Different Process Models: A Survey Report

The result of the survey carried out by Dr. Jon Holt [3], related to current practice in software engineering reveals the percentage of usage of different types of software development lifecycle models (SDLC) in practice. The result is shown below in *Figure 1*. Although different organizations do use different lifecycle models, but from the above data it is clear that *a large part of industries (22%) do not use any lifecycle model at all!* The BIG question here is why these organizations do not follow any life cycle model?

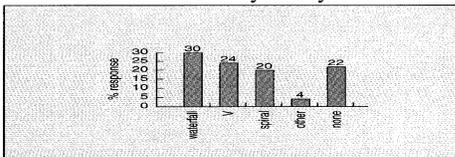

**Figure 1:** Use of different SDLC models in Practice [3]

The probable answer is, either no lifecycle model is suitable for their projects or they don't find it useful. In either case, it means the existing models lacking suitability. Hence, we need to improve the suitability of these models so that it can be used in practice.

## III. Characteristics of good Software Development Process Model

Any software development process model should have the following characteristics [2] for quality software development:

i. *The project goal reflection* i.e. the process model must reflect the project development goals.

ii. *Predictability* i.e. it must be able to forecast the out put of the project following the model prior to project completion.

iii. *Support testability and maintainability* i.e. the process model must focus on reducing the cost, effort of testing and maintenance.

iv. *Support change* i.e. the process model must handle the necessary changes.

v. *Early Defect Removal,* because the delay in error detection increases the costs to correct them.

vi. *Process improvement and feedback* i.e. each project done using the existing process model must feed information back to facilitate further process improvement.

vii. *Quantitative progress measurements* i.e. the process model at any point must give a quantitative measurement of the progress attained.

viii. *Support of process tailoring* in special situations at necessity.

## IV. Nature of Modern Software Projects

The earlier software projects were of limited scope with relatively less complexity and smaller size. In contrast, the modern software has *wider scope, higher degree of complexity and larger size with better quality, portability* and *scalability* requirements. Some times, the modern software has *to work with some existing legacy system*. Developing such system are more challenging because of the *inter-operatability* and *dependency* factors. The modern real-time systems have lots of *critical issues* such as *time and space complexity* requires to be addressed. Tremendous hardware development rate has brought us towards the system-on-chip (SOC) era. In such systems, the software has to work in coordination with the particular hardware. Developing such systems are more critical because of the hardware *constraints*. As result of advancement in network technology, more often systems are becoming *web based* and *distributed* in nature. In conclusion, the modern softwares are different in various respects from the earlier softwares.

## V. Modern Software Crisis:

*Software Crisis* may be loosely defined as the *problems associated with the software development process.* Among a lot, a few critical software crisis with modern software development are listed below [5,8]:

i. Larger *size.*
  ii. Increasing *complexity.*
  iii. Higher development *cost.*
  iv. The *delivery* challenges i.e. *late* system delivery.
  v. The *trust* challenge. How much can we trust on system operations?
  vi. *Incorrectness*: Not satisfying the client needs exactly.
  vii. Poor *quality.*
  viii. Poor *productivity.*
  ix. The *heterogeneity* Challenges i.e. inter-system coordination problem.
  x. Demand of *reusability.*
  xi. *Modularity*.
  xii. *Maintainability*.
  xiii. *Integration* problem.
  xiv. *Scalability*.
  xv. *Portability*.
  xvi. *Change* Management.
  xvii. *Risks* associated with software development.

## VI. Trends in Modern Software Development

Recently, lots of new approaches are being used at practice to overcome the modern software crisis. Some recent trends in modern software developments are listed below:
  i. Component based software development.
  ii. Software reuse
  iii. Aspect oriented software development.
  iv. Service oriented software development.
  v. Multi-Tired Software Design.
  vi. Object Oriented Software Development.
  vii. Standards practices.
  viii. Use of CASE tools.

## VII. Reasons for failure of Traditional SDLC Models: The Shortcomings

After analysing the existing SDLC models, the shortcoming of these models may be broadly summarised as follows:
  i. *Non-Involvement of the client* over the entire project development.
  ii. *Lack of better understanding* of the system requirements.
  iii. *Lack of communications* among the team members.
  iv. *Lack of project management* controls over the entire development period.
  v. *Overlooking verification* activity
  vi. *Insufficient documentations*.
  vii. *Lack of configuration management*.
  viii. *Non importance to component based* software development and
  ix. *Poor support of component reusability*.

Directly or indirectly, the above reasons are the real causes of the various software crises. I have tried to address these causes of software crisis in my proposed model discussed shortly.

## VIII. Need of Modified Process Model

Although, tailored traditional software development process models are being used since a long time, but these are not good enough at practice. Hence, we are in search of a new software development process model that will adopt and encourage these modern practices. In the forth-coming section a rather novel software development process model-BRIDGE, is proposed and discussed that attempts to encourage the modern software development trends. As well said by David Norton, research director at Gartner "I do not feel waterfall development was bad. It's given us a lot of software over the last 30 years, but I think its time is up"[1].

## IX. BRIDGE: The Proposed Model for Modern Software Development Process

After analysing the importance of all the recent software development trends, at attempt is taken to develop a rather new and novel software development process model that adopts the modern software development trends and practices. The so named BRIDGE model is the result of such an attempt, which is elaborated over the following sections. The schematic diagram of the BRIDGE model is given in *Figure 2*.

**A. BRIDGE Process Model Description:**

Unlike the other process models, the BRIDGE model consists of several phases with distinguished objectives that are discussed in the following section briefly:

**i. Phase1: Requirement Analysis, Verification and Specification**

The *objective* of this phase is to *identify the exact requirements* from the client using different techniques and to specify them in a document for future use after verification. During *requirement gathering,* the analyst extracts the system requirements from the client. In practice, it is really a tough job for the analyst to extract the requirements from the client, as the clients are unable to identify and express the exact requirements prior experiencing the system practically. The gathered requirements required to be analyzed for removing the redundancy, incompleteness, inconsistencies, anomalies etc. This phase is often called the *requirement analysis phase.* Finally, the verified requirements are to be *specified* in a document called *Software Requirements Specification (SRS)* and stored for future use. This phase is often called *requirement specification* phase. This SRS document may serve as the agreement document between the client and the company and becomes the baseline for proceeding to the next phase.

**ii. Phase2: Feasibility Analysis, Risk Analysis, Verification and Specification**

The *objective* of this phase is to *analyze the suitability* of the project in respect to different project attributes to check the different suitability aspects among the alternatives. After carrying out the analysis, the optimal solution is selected. At this stage the project cost estimation has to carry out. The different feasibility i.e. *economic feasibility*, *technical feasibility*, *operational feasibility* has to carry out to manage the different system constraints. Some times, the result of the different feasibility analysis may contradict. In such cases, necessary changes, modification and/or negotiation may have to do in the project upon consulting the client if the project is not cancelled. Finally, after verification the result of the feasibility analysis has to be specified in a document called *feasibility report and to be kept for future reference.* Beside feasibility analysis, at this phase the different project risks have *to be identified, analyzed* and specified in the *risk specification document.*

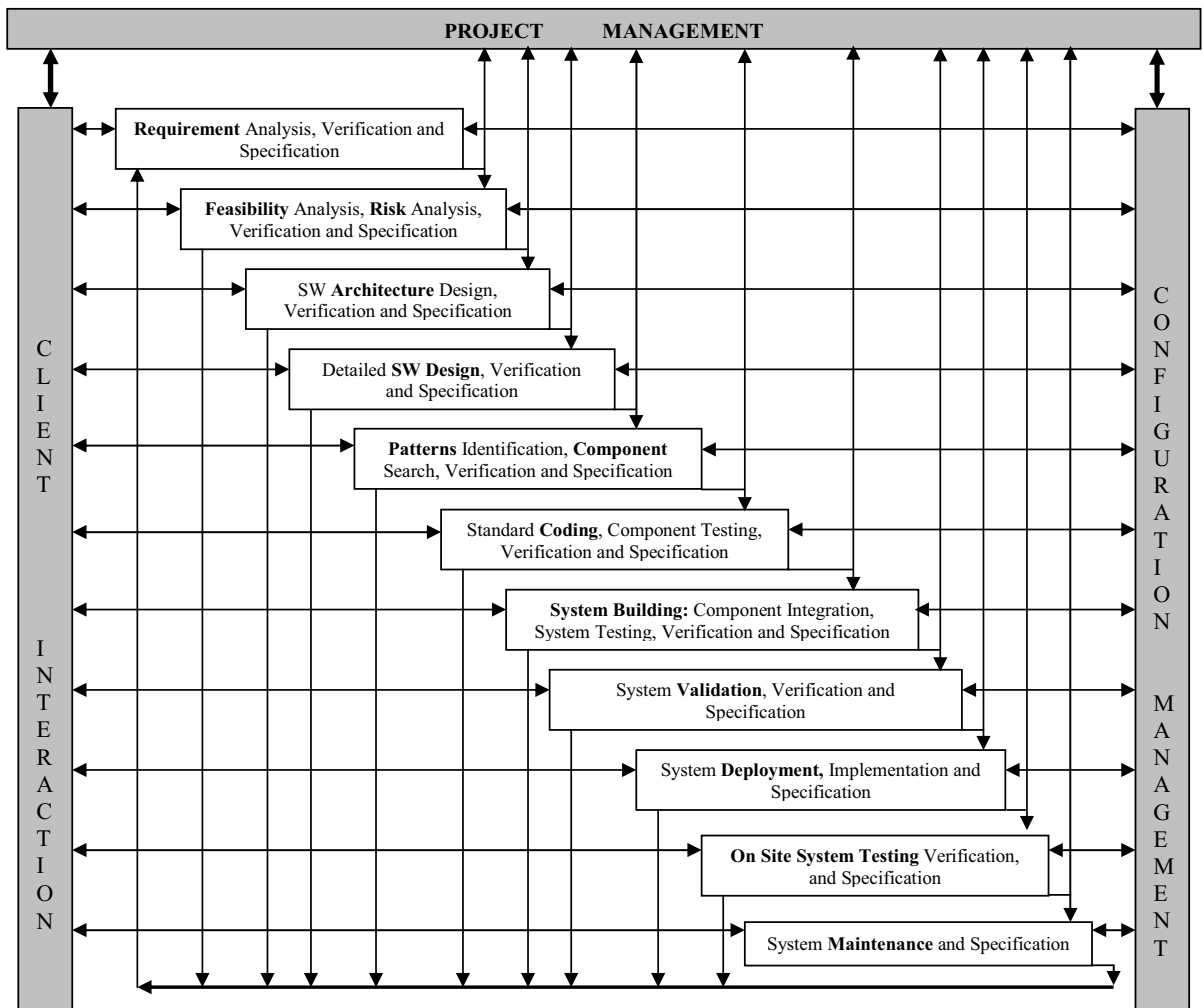

**Figure 2: BRIDGE Software Development Process Model**

### iii. Phase 3: Software Architecture Design, Verification and Specification

Once the project is confirmed, we must design the software architecture. Software architecture design is a *high-level design activity* and relatively a recent trend in industries after understanding its importance. We may consider software architecture as *abstract design* of the complete system. The *objective* of software architecture design is to *identify the sub-systems, building blocks or the components* of the system along with their *communication interfaces* expressing their *external behavior* to improve the project understandability and to communicate with the different stakeholders. The architecture design should reflect the functional requirements specified in the SRS document. Once the software architecture is designed, the architecture design must be *verified* to check whether it conforms the system requirements correctly or not. The verified software architecture design is specified in a *software architecture design document (SADD)*. It must be clear that implementational issues are not considered while designing the software architecture.

### iv. Phase 4: Detailed Software Design, Verification and Specification

In this phase, the *detailed design* of the system has to be prepared conforming the software architecture designed during the last phase. Software design is basically a *low-level design* activity *keeping the implementational issues in mind*. The *objective* of this phase is to prepare the *modular design* of the system that can be *directly implemented using some programming language*. The *data structure and algorithms* are also to be developed in this phase. The verified software design specified in a document named as *software design document (SDD)* that will be used in the other development activities later.

### v. Phase 5: Patterns Identification, Component Search, Verification and Specification

In general, a system consists of a set of sub-systems, so called *components*. If we analyze any problem, we may find some components common in different projects *representing some general structures of a system*. These common components are sometimes called *patterns*. The *objective* of this phase is to *identify these patterns*. But, to use these pre-developed components efficiently in our system, the *system must be designed keeping this objective* in mind and the designer should be well aware of the available components in the component library. From the architecture design, we must be able *to identify the components* and then it must be *searched* in the component library to find a suitable *component match*. Before moving to the next phase, we must verify the current phase properly and specifying in a document called *component specification document (CSD)* for future use.

### vi. Phase 6: Standard Coding, Unit Testing, Verification and Specification

All the components identified during the last phase may not be available in the component library. The *objective* of this phase is *to write program code for the unmatched components*. Often, a few unmatched components may work as desired just with a suitable added interface. In those cases, the benefit analysis must be done to take the decision whether to develop the interface only or the unmatched components from the scratch. *The unmatched modules must be coded properly following the standard coding guidelines and practices* laid down by the organization itself or the available standard conventions as per the organization interest. These newly developed components must be tested thoroughly since these components are going to be used in several systems at different times. Such testing is called *unit or component testing*. The components taken from the component library together with the newly developed components should be sufficient enough to build the whole system. The newly developed components may be added in the component library for future use if it looks justifiable. After verifying and specifying the phase properly, next phase can be started.

### vii. Phase 7: System Building: Component Integration, System Testing, Verification and Specification

Once all the individual *components* are gathered, it's the time to *integrate* these to build the whole system preferably following the bottom up approach. Hence, the *objective* of this phase is *to build the whole system by integrating all the components*. However, it is not necessary that, after integrating the pre-tested components successfully, the integrated system will work correctly. Various types of problems such as type mismatch, number of parameter mismatch, return type mismatch etc. may arise. Hence, there is a need to test the integrated system at different level of integration. This is called *integration testing*. Now, the complete build-up system has to be tested thoroughly using the different testing techniques to check the correctness of system functionality. The testing at this topmost level is termed as *system testing*. After performing the different testing, the corresponding *test report* has to be prepared for use during system validation and maintenance activities. Finally, the phase verification is to be carried out prior moving to the next phase.

### viii. Phase 8: System Validation, Verification and Specification

Merry successful verification of the system doesn't ensure the fulfilling of all client requirements! By successful verification of the system, we can only ensure that whatever the functions are implemented in the developed system do work correctly, but does it mean that, all the function required by client are implemented in the system? **No**. The *objective* of this phase is *to check* whether *all the functional requirements* as specified in the SRS document specified by the client *are exactly included the system or not*. There must be one to one correspondence between the functions in the SRS document and function supported by the system. Performing this activity is called *system validation*. Not only the system functionality but also the quality of the system has to be validated. Unlike the other phases, at the end, this phase to be verified and the out come of the system validation activity are to be specified in a document called *validation report* and stored for further use.

### ix. Phase 9: System Deployment, Implementation and Specification

Once the system is validated, now it's the time to deliver the system to the client and implement the system at client site. Again, some more changes may be required to accommodate and adjust for proper functioning of the system. *Delivering the system to client should not be taken as a formality! Ultimately you- the developers are not going to use the system, but the users definitely.* Until the users are not able to use the system effectively and efficiently, developing the system remains purposeless. We must facilitate the user to understand and feel comfortable with the system at use. There are basically two tools for this purpose. First, the *documentations* and second, *training*. Necessary training has to be provided to all different categories of users within their operational scope. The user refers to the documents to solve problems at any point of time during the system use. The *objective* of this phase is *to deliver, implement the system at client's work-site and train the users, if necessary.* After verification of the phase necessary documents are to be prepared and retain.

### x. Phase 10: On Site System Testing Verification, and Specification

Although, system testing is completed prior to system implementation, but due to different environmental changes and other reasons, the system may not function correctly at the work-site. Hence, after implementation, the system needs to be tested at work site too. This testing is called **on-site system testing**. The *objective* of this phase is to *check the system performance at work-site*. Finally, the *on-site system testing report* has to be prepared and to be retain after the phase verification. At this point, the current system is at work.

### xi. Phase 11: System Maintenance, Verification and Specification

Merry successful system implementation and functioning is not the end job. There is a well saying that *no software is correct at all.* Moreover, Lehman's first law related to software says, *"Software product must change continually or become progressively less useful"* [5]. Software Maintenance denotes any changes made to a software product after it has been delivered to the client. Maintenance is a continuous process over the software life cycle. The *objective* of this phase is *to provide the post delivery services to the system for its desirable functioning.* Maintenance support is to be provided to retain and improve the system quality over its lifetime. The maintenance may be of different types i.e. *corrective maintenance, adaptive maintenance, perfective maintenance and preventive maintenance* [5,6]. Finally the *maintenance report* is to be made periodically and kept for future reference.

It should be clear that deliverables from any phase might be given as input to the other phases if needed.

### xii. Phase 12: Configuration Management

*2009 IEEE International Advance Computing Conference (IACC 2009)* 497

As we have seen, *system requirements always change during system development and use*. Accordingly, these changes have to be made in associated documents and dependable. Finally, these changes are *to be incorporated into new version of the system*. Hence, it is clear that, the deliverables from different phases are to be maintained for future use. As we have seen over the past discussion that, *from each phase different documents are produced and need to be kept properly for future use*. The patterns are even to be kept in such a way that, on demand we must be able to identify, search, and locate all these components for adaptation. Hence, we need to have an efficient *document and component keeping system*. If there is no proper management and control over these changeable particulars, then it is very tough to incorporate these necessary changes in the following version of the system. *The means by which the process of software development and maintenance is controlled is called configuration management*. The **objective** of the configuration management is the development of procedures and standards for cost effective managing and controlling the changes in the evolving software system aiming *to keep track of all the important* deliverables obtained from different phases.

### xiii. Phase 13: Project Management

Unlike the configuration management activity, the project management activity has to be carried out in parallel with all the other software development phases. While developing software, we need to carryout some management activities that are part of project management. The **objective** of this phase is to *perform the project management activities including project planning, monitoring, controlling, directing, motivating and coordinating*.

### X. Analysis of the BRIDGE Model

The in-depth study of the BRIDGE model discloses a lot of information that may be used to analyze the model. These are briefly discussed below:

**A. Findings from the Study of BRIDGE Model:**

The findings from the BRIDGE model are listed below:
  ix.  It *involves the client* over the entire development life cycle activities.
   x.  It keeps continuous *communication with the project management* team.
  xi.  It explicit *verification of individual* phases.
 xii.  Separate software *architecture design* phase.
xiii.  Separate *system deployment* phase.
 xiv.  Separate *on-site system testing phase*.
  xv.  Supports *components based* software development.
 xvi.  It emphasizes on *standard coding*.
xvii.  It considers *configuration management* as a separate activity.
xviii. It forces to *specify* all the phase deliverables.
 xix.  It explicitly instructs to *validate the system*.

**B. Impact analysis of findings from BRIDGE Model Study**

In this section, impacts of the findings from BRIDGE model studies on the project goal are analysed distinctly.

*i. Impact of continuous client involvement:*
It is experienced that, as the system is more studied and analyzed over the time, the client specifies more new requirements. Satisfying these requirements, *client satisfaction* and *software quality* are improved with great impact on both *project* and *organizational goal*. Moreover, involving the client over the entire SDLC *project risks can be alleviate* up to a significant extent. By means of continuous client involvement, this model can embed the *prototyping paradigm of software development*.

*ii. Impact of continuous project management team involvement:*
The impact of involving the project management team over the SDLC model may facilitate *effective project management activities* such as *project planning, progress monitoring, project controlling*, r*isk management, Motivation* and individual *performance analysis* used for organizational and personal appraisal.

*iii. Impact of explicit verification activity:*
By verifying the individual phases indirectly the *phase entry and exit criterion* may be satisfied which reduces the error occurrence rate in the later phases. This may even overcome the well-known *99% complete syndrome problem*. Verification helps in *early error detection and correction reducing total development cost having direct impact on software testing, quality control and timely product delivery*.

*iv. Impact of software architecture design:*
Software architecture is the key framework better project understanding and communication with the various stakeholders. Software architecture has a profound influence on organization functioning and structure [4]. Designing the software architecture has the direct impact on the software quality attributes such as *performance, security, safety, availability, maintainability, scalability, productivity, cost, effort and timely product delivery*.

*v. Impact of separate system deployment phase:*
It directly maps the *environmental view supported in UML*. There is a very poor practice of considering system delivery as just a formality. Proper training must be given to the users for efficient and effective system use. More over it helps to handle all software crisis related to product deployment improving the software quality.

*vi. Impact of separate on-site system testing phase:*
The on-site testing helps to improve system quality and client satisfaction reflecting the long-term goal of the Organization.

*vii. Impact of component based software design:*
The component based software design helps in achieving better *software maintainability, reusability, productivity and quality reducing total development cost and effort*.

*viii. Impact of following standard coding:*
Following standard coding practices and conventions have remarkable impact on *better understanding* of the code written by others *reducing efforts in error isolation and system testing improving the maintainability, quality of the software. It does encourage good programming practices*.

*ix. Impact of configuration management activity:*
Configuration management activities improve different documents and components management. It does facilitate component repository and reusability reducing total development cost and efforts improving the software quality and increasing organizational assets simultaneously.

*x. Impact of document specification:*
The different specified documents facilities *better system understanding* leading to *ease error handling*. These are the *means of communication* among teammates and stakeholders. It helps in *reduction of testing and maintenance efforts*.

### xi. Impact of system validation:
System validation ensures correct system functionality by error detection achieving the goal of better quality software development. Finally, it increases degree of client satisfaction attaining long-term project and organizational goal.

## XI. Validating the BRIDGE Model in Support of Goodness Criterion

The proposed BRIDGE model does satisfy almost all the goodness criterion [2] of a good software development process. In this section, I discuss the supporting issues for validating this model against the individual goodness criteria.

### i. Support towards project goal reflection
As per the definition of software engineering given by Stephen Schach [7], the goals of software project are:
- a. Developing *quality* software.
- b. Developing the software *within budget*.
- c. Delivery of the software *within time*.
- d. *Satisfying customer* requirements.

By focusing on the phase verification and validation activities, and recommending software testing at different levels, this model reflects the goal of developing *quality software*. Again, specially performing economic feasibility analysis and involving the management over the process, the model reflects the goal of developing the software *within budget*. On stream, by involving the project management team over the entire process development model, this model puts focus on proper management control to follow the time constraints on the project development. Finally, by means of client involvement over the complete software development process, the BRIDGE model achieves the goal of *customer satisfaction*.

### ii. Support of Predictability
The software architecture is the best document to predict the different project parameters. Having a separate software architecture phase and risk analysis, this model achieves the predictability criteria.

### iii. Support of testability and maintainability
Emphasizing on component based software development and component reusability concept, this model highlights the testability criteria. In addition, designing the software architecture gives the foundation for meeting maintainability criteria with a separate phase related to software maintenance.

### iv. Support towards change
Designing software architecture and by supporting maintainability, this model achieves the change management criteria directly with consistent support from configuration management.

### v. Support of early defect removal
By involving the customer over the entire development process, it is possible to detect errors at earliest and performing verification activity following each phase ensures early defect detection and removal.

### vi. Support of process improvement and feedback
During the configuration management activities, all the prepared documents and reports are stored. Project completion analysis report with the available documents and the reports from configuration database, can be used to judge and identify the activities needing process improvement and applying the same in the next project. Customer comments and recommendations can be used as the feedback for further process improvements.

### vii. Support of Quantitative Progress Measurement
Directly, each phase indicates a milestone towards the project completion. All the deliverables from various phases of this process model can be used to measure the progress of the work completed.

### viii. Support of Process Tailoring
Since the process activities are decomposed in several phases, at necessity, more than one phase can be combined and any phase can be further decomposed into sub phases or even might be dropped depending on the project characteristics. Hence, it may be concluded that, the BRIDGE model satisfies all the desired characteristics of a good software process model.

## XII. Suitability of the BRIDGE Model
This model can be used to both simple systems as well as complex systems. It supports the object oriented, component based software development paradigm. By process tailoring, this model also can be applied to develop any software projects that are directly unfit to the actual model. Hence, the suitability of the BRIDGE model for any modern software development is justified and may be recommended for any kind of software project development.

## XIII. Limitations of the BRIDGE Model
Along with the strong suitability, this model has some limitations as pointed down below:
- a. Non-considering the implementational issues.
- b. Abstracts the different techniques to be used in different phases.
- c. Required to be validated by industrial practice.
- d. It doesn't consider professionals skill level.
- e. The BRIDGE model seems to be complex.

## XIV. Naming Significance: BRIDGE
The schematic diagram of the proposed model looks like a bridge. In a bridge, the entire load is on the bridge floor, but this load has distributed over all the pillars for its survivals. Directly the project pressure is on the "Project Management" and this pressure has to be distributed over "Client Interaction", "Configuration management " and other the phases indirectly- the pillars of the model. Keeping this point of view the name, BRIDGE, is given and justified.

## XV. Conclusion
After the complete analysis, it can be conclude that if the BRIDGE model is followed to any software project development, most of the software crisis may be overcome up to great extent delivering the fully functional system with better quality within time and budget achieving the true goal of any software project development.